\documentclass[groupedaddress,twocolumn,prl]{revtex4}
\usepackage{epsfig,amsfonts,amsmath}
\newcommand{\be}{\begin{equation}}
\newcommand{\ee}{\end{equation}}
\newcommand{\ba}{\begin{eqnarray}}
\newcommand{\ea}{\end{eqnarray}}

\def\bea{\begin{eqnarray}}
\def\eea{\end{eqnarray}}
\def\ben{\begin{eqnarray*}}
\def\een{\end{eqnarray*}}

\def\>{\rangle}
\def\<{\langle}
\def\l{\left}
\def\r{\right}
\def\Sr{^{88}\text{Sr}^+}

\newcommand{\nc}{\newcommand}
\nc{\cN}{{\cal N}}

\def\l{\lambda}

\def\r{\rho}

\def\be{\begin{equation}}
\def\ee{\end{equation}}
\def\bea{\begin{eqnarray}}
\def\eea{\end{eqnarray}}
\def\ben{\begin{eqnarray*}}
\def\een{\end{eqnarray*}}

\def\>{\rangle}
\def\<{\langle}

\def\l{\left}
\def\r{\right}

\newcommand{\fig}[1]{Fig.~\ref{fig:#1}}

\newcommand{\ord}[1]{\mathcal{O}(#1)}

\def\etal{\emph{et al.}}

\begin{document}

\title{A cryogenic surface-electrode elliptical ion trap for quantum simulation}
\author{Robert J. Clark, Ziliang Lin, Kenan S. Diab, and Isaac L. Chuang}
\affiliation{Center for Ultracold Atoms, Massachusetts Institute of Technology, Cambridge, MA, 02139}

\begin{abstract}

Two-dimensional crystals of trapped ions are a promising system with which to implement quantum simulations of challenging problems such as spin frustration. Here, we present a design for a surface-electrode elliptical ion trap which produces a 2-D ion crystal and is amenable to microfabrication, which would enable higher simulated coupling rates, as well as interactions based on magnetic forces generated by on-chip currents. Working in an 11~K cryogenic environment, we experimentally verify to within 5~\% a numerical model of the structure of ion crystals in the trap. We also explore the possibility of implementing quantum simulation using magnetic forces, and calculate $J$-coupling rates on the order of $10^3$ s$^{-1}$ for an ion crystal height of 10~$\mu$m, using a current of 1~A.

\end{abstract}

\pacs{} 

\maketitle

Quantum simulation is a technique whereby one may calculate properties of one quantum-mechanical system using a different, more controllable quantum system \cite{Feynman:82}. It has been shown, in principle, to hold great potential for solving problems in many-body quantum physics that are believed to be intractable on classical computers \cite{Lloyd:96,Abrams:97,Wu:02,Brown:06}. Recently, a particular interest has arisen in the simulation of quantum spin models using trapped ions \cite{Porras:04}, and experiments on two and three ions confined in linear traps have been demonstrated \cite{Schaetz:08,Kim:10}. Direct analog quantum simulation of spin frustration could be performed using a two-dimensional (2-D) array of trapped ions. In addition, such arrays have been proposed for more general quantum computations \cite{Porras:06,Cirac:00}. 

Prior work toward developing 2-D arrays of ions has fallen within two paradigms: arrays of individual ion traps, and Coulomb crystals within a single trap region. Arrays of individual traps have been suggested as a way to implement simulations based on either optical forces \cite{Porras:04} or magnetic gradient forces \cite{Chiaverini:08}. However, for realistic experimental parameters, both the motional and simulated interaction rates appear to be well below the expected decoherence rates in such ion traps \cite{Clark:09}, even when scaling to microscopic ion-ion distances. Despite recent advances in the design of arrays of surface-electrode ion traps \cite{Schmied:09}, this scaling is likely to be problematic for some simulation protocols. Arrays of Coulomb crystals within a single trap, as realized in Refs.~\cite{Block:00} and \cite{Itano:98}, bring other challenges, among them the $\vec{E} \times \vec{B}$ rotation of the ion crystal in Penning traps and rf heating due to micromotion in Paul traps. Recent theoretical work suggests that quantum simulation may be performed using 2-D crystals of $\ord{100}$ ions in linear ion traps \cite{Buluta:08}, even with the effects of rf and background gas heating. Although such traps may suffice for quantum simulation using optical forces, it is easier to achieve high magnetic field gradients if the trap scale can be decreased. Magnetic interactions provide the additional advantages that errors due to spontaneous laser scatter are removed \cite{Chiaverini:08,Mintert:01} and that magnetic field gradients may be made uniform over the entire trap volume, enabling convenient global operations and mitigating errors due to micromotion.

\begin{figure}[h]
\begin{center}
\includegraphics[width=.4\textwidth]{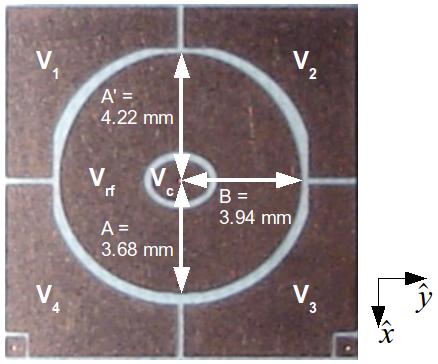}
\end{center} 
\caption{Photograph of the elliptical ion trap. The rf electrode is labeled as $V_{\text{rf}}$, while the others are dc electrodes. The rf electrode is approximately elliptical; however, in order to tilt the trap principal axes and improve Doppler cooling, the semiminor axis of the electrode is stretched along one direction. The semimajor axis is labeled B in the figure, while the semiminor axes are labeled A and A$^{\prime}$. The semimajor and semiminor axes of the center electrode are 0.94~mm and 0.71~mm, respectively. The coordinate system is defined in the figure; the $\hat{z}$ axis points out of the page.}
\label{fig:fig1}
\end{figure}

In this Letter, we present the design and proof-of-principle demonstration of a Paul trap that generates a 2-D crystal of trapped ions and that is amenable to microfabrication. Our trap is a surface-electrode ring trap, similar to the trap presented in Ref.~\cite{THKim:10}, but with the usual cylindrical symmetry broken in two ways. The first is that the center ground and rf ring electrodes are based on ellipses rather than on circles, similar to the elliptical trap invented by DeVoe \cite{DeVoe:98}, in order to break the degeneracy of the in-plane motional modes and give the ion crystal a unique orientation in space. The second is that the rf ring electrode is made wider on one side than the other, to tilt the ion crystal with respect to the plane of the ion trap and improve Doppler cooling. We first present a numerical model of the trap, along with experimental measurements which verify the model. We then turn to the question of what conditions must be met for a quantum simulation to be performed within such a trap. To that end, we explore theoretically how quantum simulations may be done using magnetic field gradients in a scaled-down version of the elliptical trap. 

The trap is a printed circuit board consisting of copper electrodes on a low-loss Rogers 4350B substrate, the same manufacturing process that was used in Refs.~\cite{Brown:07} and \cite{Leibrandt:07}. A photograph of the trap is presented in \fig{fig1}. The trap was numerically modeled using the Charged Particle Optics (CPO) boundary-element software package. Typical secular frequencies for an rf drive frequency of $\Omega/(2\pi) = 3.5$~MHz and an rf amplitude of $V_{rf} = 150$~V are $\nu_x = 124$~kHz, $\nu_y = 110$~kHz, and $\nu_z = 234$~kHz. Confinement may achieved by applying an rf voltage and grounding all dc electrodes. 

A cryogenic environment was used for our experiments, which offers the advantages of suppressed electric field noise \cite{Labaziewicz:08} and a rapid pumpdown to ultra-high vacuum pressures. The trap was mounted in a closed-cycle cryostat (ARS model DE-210 with DMX-20 anti-vibration interface), similar to one described in Ref.~\cite{Antohi:09}. The electrode labeled $V_3$ was thermally anchored to the baseplate, which typically reached a temperature of 11~K. The vacuum pressure was upper-bounded by reading the current on a 20~l/s ion pump that was connected to the room-temperature vacuum housing. The vacuum pressure at room temperature was at most $5 \times 10^{-10}$~torr, and within the cryogenic environment was expected to be much lower. $\Sr$ ions were loaded by photoionization of a neutral strontium beam and Doppler cooled on the 422~nm 5S$_{1/2}$ $\rightarrow$ 5P$_{1/2}$ transition with a 1092~nm repumper. Typical powers were 50~$\mu$W of 422~nm and 100~$\mu$W of 1092~nm with beam waists of $\approx 50$~$\mu$m. 

For the measurements presented here, an rf amplitude of $V_{rf} = 150$~V at $\Omega/(2\pi) = 3.5$~MHz was used.  Stray fields were compensated by adjusting the dc voltages such that a cloud of ions remains stationary when $V_{rf}$ is varied \cite{Berkeland:98}. The final set of voltages used was $V_1 = -3.90$~V, $V_2 = 1.56$~V, $V_3 = V_4 = 0$~V, and $V_c = -2.62$~V. Secular frequencies were then measured by applying a low-amplitude resonant signal to one of the dc electrodes, and recording the dip in the ion cloud's fluorescence as the ions were resonantly heated. For the above conditions, we measure $\nu_x = 177$~kHz, $\nu_y = 141$~kHz, and $\nu_z = 414$~kHz, with an error $\approx 2$~kHz on each. These frequencies differ from those listed above due to the fact that the dc voltages used for compensation also alter the curvature of the trapping potential. 

\begin{figure}
\begin{center}
  \begin{minipage}{0.23\textwidth}
   \includegraphics[width=\textwidth]{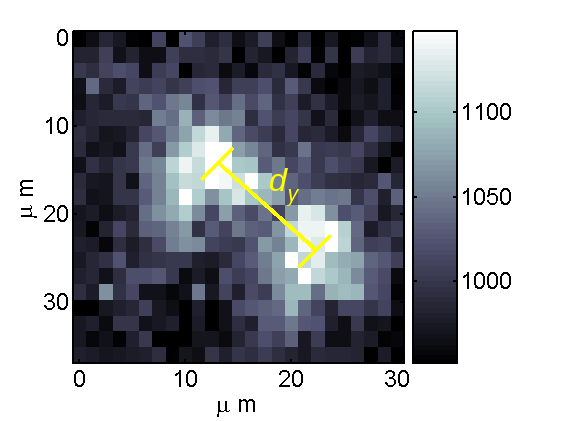}
  \end{minipage}
  \begin{minipage}{0.23\textwidth}
   \includegraphics[width=\textwidth]{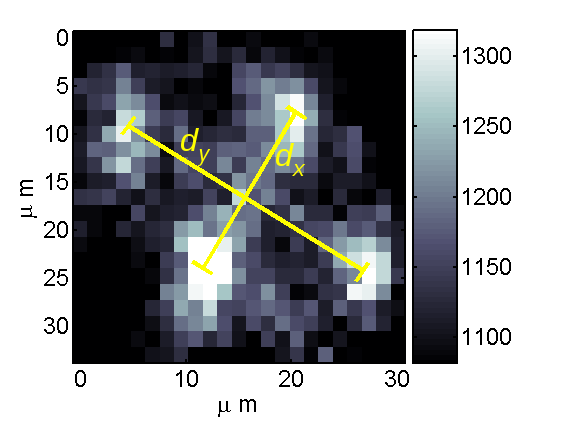}
  \end{minipage}
\end{center} 
\caption{Images of crystals of two and four ions in the elliptical trap. The ion-ion spacings $d_y$ and $d_x$ are along $\hat{y}$ and $\hat{x}$, respectively. Since the separation of two ions may be calculated analytically, the two-ion crystal is used to calibrate the magnification of the optics system. For two ions, $d_y = 16.5$~$\mu$m. For four ions, $d_y = 28 \pm 3$~$\mu$m and $d_x = 17 \pm 3$~$\mu$m, compared to the calculated values of $d_y = 29.6$~$\mu$m and $d_x = 17.1$~$\mu$m. Note that $d_x \neq d_y$ due to the eccentricity of the trap.} 
\label{fig:fig2}
\end{figure}

The structure of crystals in the trap is calculated by numerically minimizing the potential energy of a set of $N$ ions in a harmonic potential characterized by the three secular frequencies. For sufficiently small $N$, the crystal extends only along $\hat{x}$ and $\hat{y}$. For $\nu_x \approx \nu_y$, the condition for a planar crystal may be written as $\nu_z > \sqrt{70 N \nu_x^4 / \pi^3}$, as noted in Ref.~\cite{Buluta:09}. For $\nu_x \neq \nu_y$, such a condition may be computed numerically. \fig{fig2} contains images of crystals of two and four ions, along with their measured separations. We find that for the measured secular frequencies, the agreement is quite good. Thus, our model of the elliptical trap potential and crystal structure therein is validated, at least for small numbers of ions. A limited 422~nm laser stability, along with the possibility of rf heating, are possible technical explanations for why larger ion crystals were not obtained. Larger crystals were later observed in a similar ion trap (with circular geometry) in the same cryogenic apparatus \cite{THKim:10}.

This trap design provides a unique platform for quantum simulation, in that wires that produce magnetic field gradients may be integrated into or beneath the ground electrode ($V_c$). Although quantum operations based on oscillating magnetic fields have been proposed \cite{Ospelkaus:08}, we concentrate here on static field gradients. For a given configuration of current-carrying wires, the magnetic field $\vec{B}$ above the trap is  calculated by direct numerical integration of the Biot-Savart law. The state-dependent force $\vec{F}$ on a trapped ion is given by $\vec{F} = - \nabla \l ( \vec{\mu} \cdot \vec{B} \r )$, where $\vec{\mu}$ is the magnetic moment of the ion. A qubit may be formed from either a Zeeman-split electronic ground state or a pair of hyperfine levels. 

\begin{figure}
\begin{center}
  \begin{minipage}{0.17\textwidth}
   \includegraphics[height=.8\textwidth]{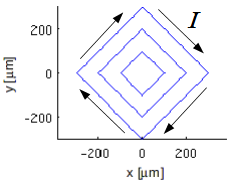}
  \end{minipage}
  \begin{minipage}{0.30\textwidth}
   \begin{tabular}{| c | c | c | c |}
   \hline
   $H$ ($\mu$m) & $d$ ($\mu$m) & $F$ (N) & $J$ (s$^{-1}$) \\ 
   \hline
   10 & 0.9 & $3\times 10^{-20}$ & 600 \\
   50 & 2.7 & $2\times 10^{-21}$ & 30 \\
   100 & 4.3 & $6\times 10^{-22}$ & 5 \\
   \hline 
   \end{tabular}
  \end{minipage}
\end{center} 
\caption{Left: Example wire configuration in the plane of an elliptical ion trap. Right: A table of some values of the simulated coupling rate $J$ as a function of the ion height $H$ above the trap surface, for $^{88}\mathrm{Sr}^+$. The ion height is varied by scaling down the entire trap structure, including the current-carrying wires, and the corresponding approximate ion-ion spacing $d$ is given for reference. We assume a current of 1~A through each wire. The secular frequencies are scaled up in inverse proportion to the trap scale. The quadratic dependence of $J$ on $F$ indicates that these values of $J$ could be improved by a modest increase in the current or in the number of concentric wire loops.} 
\label{fig:fig3}
\end{figure}

As an example of how a quantum simulation may be implemented using this system, we calculate the magnetic fields for three concentric square wires embedded within the center ellipse. Each is assumed to carry a current of 1~A, on par with the currents used in neutral atom chip traps \cite{Haensel:01}. The wire configuration is depicted in \fig{fig3}. Using these calculated force values, and assuming the orientation of the atomic magnetic moment is along $\hat{y}$ (as set by an external bias field), one may obtain an order of magnitude estimate for the simulated interaction rate of a quantum simulation. We choose to calculate the effective $J$-coupling rate for a simulation of quantum spin models; according to Ref.~\cite{Porras:04}, and for two ions undergoing an antiferromagnetic simulated coupling, this rate is given by $\hbar J = \kappa e_c^2 F^2/(64 \pi^5 \epsilon_0 m^2 \nu^4 d^3)$, where $\kappa$ is a constant of order unity, $F$ is the magnitude of the state-dependent force, $m$ is the ion mass, $\nu$ is the relevant secular frequency, $\epsilon_0$ is the permittivity of free space, $e_c$ is the ion charge, and $d$ is the mean distance between neighboring ions. Some calculated values of $F$ and $J$ as a function of the overall trap scale are given in \fig{fig3}, showing that a trap scale on the order of tens of microns can produce $\ord{10^3~\mathrm{s}^{-1}}$ coupling rates. 

Although sufficient interaction rates for the simulation of quantum spin models are theoretically possible in our trap, there are some other effects that may reduce the fidelity of a quantum simulation that must be addressed. These include heating of the ion crystal, micromotion, and variations in the distance between ions. The heating rates of planar ion crystals in linear ion traps have been calculated in Ref.~\cite{Buluta:09}; the authors predict that for up to $\ord{100}$ ions, high-fidelity quantum simulations should be possible despite modest heating. In a 2-D crystal, some micromotion will be present, but micromotion represents a breathing-mode oscillation, in which the relative micromotion amplitude between nearest-neighbors is much less than the total amplitude of micromotion. Furthermore, magnetic gradient-based forces vary much less rapidly in space than optical forces, which mitigates the effects of micromotion. Finally, we note that the ion-ion spacing is quite uniform near the center of a many-ion crystal, provided $\nu_x \approx \nu_y$. Further calculations indicate that, with appropriate dc and rf voltages, a planar crystal of $\ord{100}$ ions could be produced in our trap \cite{Clark:thesis}. Assuming perfect quantum operations, this number would in principle
be enough to outperform exact classical simulations. While the possible number of ions in a planar crystal is comparable to the state of the art before this work, our trap design offers the possibility of harnessing the benefits of microfabricated ion trap technology, including magnetic forces, for quantum simulation. 

We gratefully acknowledge assistance from P. Antohi and S. Wang, and funding from the NSF Center for Ultracold Atoms.

\end{document}